\documentclass[pre]{revtex4-2}
\usepackage{amssymb}
\usepackage{graphicx}
\usepackage{bm}
\usepackage{physics}
\newcommand{\vect}[1]{\boldsymbol{#1}}

\begin{document}

\title{Hydrodynamic efficiency limit on a Marangoni surfer} 
\author{Abdallah Daddi-Moussa-Ider}
\affiliation{Max Planck Institute for Dynamics and Self-Organization (MPIDS), 37077 G\"ottingen, Germany}
\affiliation{School of Mathematics and Statistics, The Open University, Walton Hall, Milton Keynes MK7 6AA, United Kingdom}
\author{Ramin Golestanian}
\affiliation{Max Planck Institute for Dynamics and Self-Organization (MPIDS), 37077 G\"ottingen, Germany}
\affiliation{Rudolf Peierls Centre for Theoretical Physics, University of Oxford, Oxford OX1 3PU, United Kingdom}
\author{Andrej Vilfan}
\email{andrej.vilfan@ds.mpg.de}
\affiliation{Max Planck Institute for Dynamics and Self-Organization (MPIDS), 37077 G\"ottingen, Germany}
\affiliation{Jo\v{z}ef Stefan Institute, 1000 Ljubljana, Slovenia}
\date{\today}

\begin{abstract}
A Marangoni surfer is an object embedded in a gas-liquid interface, propelled by gradients in surface tension. We derive an analytical theorem for the lower bound on the viscous dissipation by a Marangoni surfer in the limit of small Reynolds and capillary numbers. The minimum dissipation can be expressed with the reciprocal difference between drag coefficients of two passive bodies of the same shape as the Marangoni surfer, one in a force-free interface and the other in an interface with surface incompressibility. The distribution of surface tension that gives the optimal propulsion is given by the surface tension of the solution for the incompressible surface and the flow is a superposition of both solutions. For a surfer taking the form of a thin circular disk, the minimum dissipation is $16\mu a V^2$, giving a Lighthill efficiency of $1/3$. This places the Marangoni surfers among the hydrodynamically most efficient microswimmers. 
\end{abstract}

\maketitle

\section{Introduction}

The Marangoni effect describes the motion of liquids due to surface tension gradients, caused, for example, by an uneven distribution of surfactants, and can be harnessed to drive the motion of a microswimmer.  The Marangoni propulsion can thereby take place on the swimmer's surface, like in the case of chemically active emulsion droplets \citep{thutupalli2011swarming, herminghaus2014interfacial,izri2014self,Maass.Bahr2016,Schmitt.Stark2016,testa2021Sustained,hokmabad2022chemotactic,michelin2023self}. The more common design, however, involves particles embedded in a flat gas-liquid interface that emit surfactants and thereby ``manipulate'' the surface tension in the interface around them. A classical example are the camphor boats, already studied by \citet{tomlinson1862ii} and \citet{rayleigh1890}.
These surfers create a surface tension gradient by asymmetric emission of the surfactant, but it is also possible for a symmetric swimmer to achieve directed propulsion through spontaneous symmetry breaking \citep{boniface2021role}. Marangoni swimmers have also been fabricated using accessible pen-drawn patterns and powered by fuel in the form of ink, enabling controlled movement and navigation~\citep{song2023pen}. Finally, some animals like the \textit{Microvelia} water striders use the Marangoni effect to slide along the surface of water \citep{Bush.Hu2006}.

For certain distributions of the surface tension, the propulsion of a thin disk-shaped Marangoni surfer has been exactly solved \citep{Lauga.Davis2012,Elfring.Squires2016, crowdy2021viscous}. Using the Lorentz reciprocal theorem \citep{masoud2014reciprocal}, the propulsion velocity of a surfer with any shape can in principle be determined provided the solution of the passive particle pulled by an external force is known. Solutions including inertia at finite Reynolds numbers have also been determined \citep{ender2021diffusive}. Moreover, two dimensional squirmer models share certain similarities with the Marangoni effect \citep{PhysRevE.90.032304}. 

The energetic efficiency of a microswimmer has been defined by \citet{lighthill1952} as the power needed to pull the swimmer through the fluid by an external force, divided by the actual power dissipated when the swimmer actively moves at the same velocity. Although Lighthill's efficiency can theoretically exceed 1 \citep{leshansky2007}, swimming microorganisms typically achieve values around $1\,\%$ \citep{osterman2011}. The theoretical efficiency limit of the 3-sphere model swimmer is also of a similar order of magnitude~\citep{Nasouri.Golestanian2019}.
The efficiency of autophoretic colloidal microswimmers is many orders of magnitude lower than that and is limited by the thickness of the boundary layer \citep{sabass2010}.
We have recently derived a minimum dissipation theorem that provides a lower bound on the dissipation by a microswimmer moving through bulk fluid, first for external dissipation alone \citep{Nasouri.Golestanian2021} and later for the combination of internal and external dissipation \citep{DaddiMoussaIder.Vilfan2023}.
The theorems share the common structure that expresses the dissipation with the reciprocal difference between the drag coefficients of two passive bodies.
For example, for the swimmer without internal dissipation, these would be two bodies of the same shape as the swimmer, one with the no-slip and one with the perfect-slip boundary \citep{Nasouri.Golestanian2021}. These solutions lead to the question whether a similar theorem can be derived for the hydrodynamic efficiency of Marangoni surfers. They differ from the previously solved dissipation problems in two main aspects: first, the propulsive force acts on an infinite plane outside the surfer and second, the force cannot be optimised freely, but needs to have the form of a surface tension gradient. In this paper, we derive such a theorem -- first in a general form and then for a thin surfer with the shape of a circular disk.

\begin{figure}
  \begin{center}
    \includegraphics[width=0.95\textwidth]{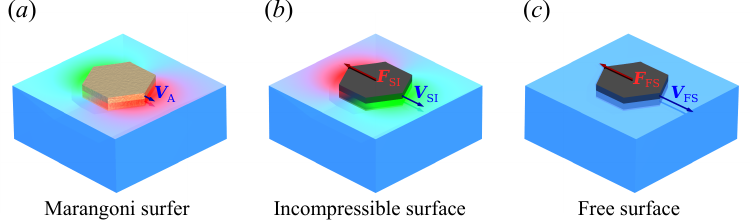}
  \end{center}
  \caption{\label{fig:1}(\textit{a}) The optimal Marangoni surfer. The surface colour indicates increased (red) or reduced (green) surface tension leading to the Marangoni effect. (\textit{b}) A passive object, pulled along the incompressible surface of a fluid with the velocity $\bm{V}_\mathrm{SI}$. The colours denote the surface tension that builds up in order to maintain the incompressibility condition. (\textit{c}) A passive object, pulled with velocity $\bm{V}_\mathrm{FS}$ in a fluid with a free surface (with a uniform surface tension).}
\end{figure}

\section{Minimum dissipation theorem for Marangoni surfers}

\subsection{Problem formulation}
In this section we derive a lower bound for the energy dissipation by a Marangoni surfer of arbitrary shape moving along the surface of a fluid at zero Reynolds number (Fig.~\ref{fig:1}a). The surfer comprises a body partially immersed in a gas-liquid interface and suspended by surface tension. The surfer can modify the surface tension in its surroundings, for example by emitting surfactants. For the purpose of this study, we disregard the dynamics of surfactants and treat the surface tension as a given. We assume that the interface remains planar, i.e., it is neither distorted by the induced flows (this corresponds to the low capillary number limit) nor by the contact angle at the surfer boundary. The surfer is force-free, driven solely by the surface tension at the contact line and by the flows induced by the Marangoni effect. The submerged part of the surfer forms a no-slip boundary with the fluid. The interface extends across the $xy$ plane, with the~$z$ direction oriented perpendicular to this plane.

The fluid motion in bulk ($z<0$) is governed by the Stokes equation along with the incompressibility condition 
\begin{subequations} \label{eq:fluid-motion-UND-incompressibility}
  \begin{eqnarray}
    -\boldsymbol{\nabla}p + \mu \boldsymbol{\nabla}^2 \bm{v} &=& \bm{0} \, , \label{eq:fluid-motion} \\
    \boldsymbol{\nabla} \cdot \bm{v} &=& 0 \, , \label{eq:incompressibility}
  \end{eqnarray}
\end{subequations}
where $\bm{v}$ and~$p$ denote the velocity and pressure fields in the fluid medium, and~$\mu$ the shear viscosity.

At the submerged part of the swimmer surface $\cal S$, not necessarily at $z=0$ if the swimmer has a non-flat shape, the fluid is subject to the no-slip boundary $\bm{v}=\bm{V}_\mathrm{A}$, where $\bm{V}_\mathrm{A}$ is the translational velocity of the swimmer ($\bm{V}_\mathrm{A}\perp \hat{\vect{e}}_z$).
The boundary condition at the gas-liquid interface located at $z=0$ implies $\hat{\vect{e}}_z \cdot \bm{v}=0$.

\begin{figure}
	\begin{center}
		\includegraphics[width=0.7\textwidth]{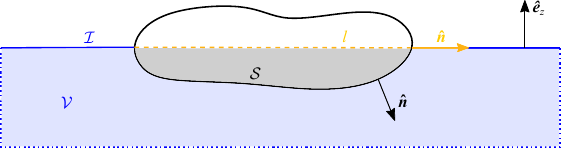}
	\end{center}
	\caption{\label{fig:2}Side view of the surfer. The fluid volume $\mathcal{V}$ is shown in blue, the gas-liquid interface $\mathcal{I}$ as solid blue line, the submerged surface of the swimmer $\mathcal{S}$ in grey and the contact line $\ell$ in yellow.}
\end{figure}

The horizontal force balance on the swimmer states $(\bm{I}-\hat{\vect{e}}_z \hat{\vect{e}}_z) \cdot \int_{\mathcal{S}} \dd S \, \boldsymbol{\sigma} \cdot \hat{\vect{n}} + \int_\ell \dd s \, \hat{\vect{n}} \gamma=0$. The first term represents the tractions exerted on the swimmer by the fluid and the second term the effect of the surface tension. Here $\hat{\vect{n}}$ denotes the surface normal pointing into the fluid and $\ell$ the contact line between the swimmer and the gas-liquid interface (Fig.~\ref{fig:2}). In the integral over $\ell$, $\hat{\vect{n}}$ denotes the in-plane normal to the contact line.
The stress tensor is determined as $\boldsymbol{\sigma}=-p \bm{I}+2\mu \bm{E}$, with the strain-rate tensor $\bm{E}=(\boldsymbol{\nabla} \bm{v} + (\boldsymbol{\nabla} \bm{v})^\top)/2$. The surface tension, reduced by its value in the surface that is not affected by the presence of the surfer, is denoted by $\gamma$. At the gas-liquid interface, the force balance states 
$(\bm{I}-\hat{\vect{e}}_z \hat{\vect{e}}_z) \cdot \boldsymbol{\sigma}\cdot \hat{\vect{e}}_z +\boldsymbol{\nabla}_\mathrm{s} \gamma=0$. Here, $\boldsymbol{\nabla}_\mathrm{s} = \hat{\vect{e}}_x \partial_x + \hat{\vect{e}}_y \partial_y$ represents the gradient within the horizontal plane. 

The total energy dissipation can either be written as the volume integral of the local dissipation rate, or the rate of work exerted by the surface tension, integrated over its surface \cite{happel1983}: 
\begin{equation}
  P=\int_{\mathcal{V}} \dd V \, 2\mu \bm{E} : \bm{E} = \int_{\mathcal{I}} \dd S \, \bm{v} \cdot \boldsymbol{\sigma}\cdot \hat{\vect{e}}_z + \int_{\mathcal{S}} \dd S \, \bm{v} \cdot \boldsymbol{\sigma}\cdot (-\hat{\vect{n}})\;.
\end{equation}  
By taking into account the no-slip boundary at $\mathcal{S}$ and the force balance on the swimmer and at the interface, the dissipation can be expressed as
\begin{equation}
  P=  \int_{\mathcal{I}} \dd S \, \bm{v} \cdot \boldsymbol{\nabla}_\mathrm{s}\gamma + \int_\ell \dd s \, \gamma \hat{\vect{n}}\cdot \bm{V}_\mathrm{A}\;.
  \label{eq:dissipation}
\end{equation}
Finally, we apply the 2D divergence theorem $\int_{\mathcal{I}} \dd S \boldsymbol{\nabla}_\mathrm{s}\cdot(\gamma \bm{v})=-\int_\ell \dd s \,  \hat{\vect{n}}\cdot (\gamma \bm{v})$ to express the dissipation as
\begin{equation}
  P= - \int_{\mathcal{I}} \dd S \, \gamma \boldsymbol{\nabla}_\mathrm{s}\cdot \bm{v} \, ,
  \label{eq:dissipation2}
\end{equation}
which represents the rate of change of the total surface energy. 

\subsection{Derivation of the theorem}
In the following we derive a theorem for a lower bound on the dissipation by a Marangoni surfer moving with the velocity $\bm{V}_\mathrm{A}$. We follow the approach that led to a minimum dissipation theorem for surface-propelled swimmers with external dissipation \citep{Nasouri.Golestanian2021} and later with combined external and internal dissipation \citep{DaddiMoussaIder.Vilfan2023}. 
Using this theorem, the flow field of an optimal nearly spherical swimmer with external dissipation has been obtained using a perturbative analytical method~\citep{DaddiMoussaIder.Golestanian2021}.

The solution consists of two major steps. The first is to find a passive minimum dissipation theorem for flows that satisfy the velocity boundary condition on an object. This passive theorem can be seen as a generalisation of the Helmholtz minimum dissipation theorem that states that among all incompressible flows that satisfy the same fixed-velocity boundary condition, the Stokes flow has the smallest dissipation \citep{guazzelli2009}. For example, for surface-driven swimmers with external dissipation, the passive problem consists of a perfect-slip body. In the second step, another passive problem is needed that also satisfies the boundary condition, but is orthogonal to the active problem in the sense that their superposition flow has a dissipation that is the sum of the dissipations of the two flows, each on its own. The minimum dissipation theorem is obtained by applying the inequality from the first condition to this superposition flow.

As a passive minimum dissipation theorem, we use a variant of the Helmholtz minimum dissipation theorem that is derived in Appendix \ref{sec:helmholtz}. The theorem states that among flows that satisfy the no-slip boundary condition on the swimmer moving with a given velocity and zero normal velocity at the fluid-gas interface, the flow without tangential tractions at the interface has the smallest dissipation (Fig.~\ref{fig:1}c). Because of the stress-free interface (free surface), we label this solution $\bm{v}_\mathrm{FS}$. If the passive body with the shape of the Marangoni surfer is moved with a velocity $\bm{V}_\mathrm{FS}$, the drag force acting on it is $\bm{F}_\mathrm{FS}= -\bm{R}_\mathrm{FS}\cdot \bm{V}_\mathrm{FS}$ with the free-surface drag coefficient $\bm{R}_\mathrm{FS}$. The total dissipation of any flow, possibly with additional surface forces, then satisfies the inequality
\begin{equation}
  \label{eq:passmdt}
  P\ge  \bm{V}_\mathrm{FS} \cdot \bm{R}_\mathrm{FS}\cdot \bm{V}_\mathrm{FS} \;.
\end{equation}
The equality is satisfied exactly when there are no additional horizontal forces beyond those pulling the passive body. In the following, we assume that the symmetries of the surfer are such that translational and rotational drag are decoupled. Then a force on the body leads to purely translational and a torque to purely rotational motion. The translational drag is then described by a symmetric $2\times2$ matrix $\bm{R}_\mathrm{FS}$. 

The second step towards the active minimum dissipation theorem requires finding a problem that is orthogonal to the active problem (Marangoni surfer) in the sense that the dissipation in the superposition of both flows is additive. We will show that this condition is fulfilled for a flow that satisfies the no-slip boundary at the swimmer body and is surrounded by a gas-liquid interface with an incompressible surface, $\boldsymbol{\nabla}_\mathrm{s} \cdot \bm{v}_\mathrm{SI}=0$ at $z=0$ (Fig.~\ref{fig:1}b). The incompressibility is achieved by a buildup of a passive surface tension $\gamma_\mathrm{SI}$, which we also define relative to the unperturbed surface. The incompressible surface occurs in the case of insoluble surfactants in the limit of an infinite Marangoni modulus \citep{Elfring.Squires2016,Manikantan.Squires2020}. The same model also represents a limiting case of the Boussinesq-Scriven model \citep{scriven1960dynamics, manikantan2017irreversible} with an incompressible ``membrane'', however without additional surface viscosity, i.e., in the limit of vanishing Boussinesq number \citep{Stone.Masoud2015}.

The superposition of the Marangoni surfer, moving with velocity $\bm{V}_\mathrm{A}$ and the passive object with incompressible surface moving with $\bm{V}_\mathrm{SI}$, has a dissipation rate that can be expressed in analogy to Eq.~(\ref{eq:dissipation}). It corresponds to the sum of the work done by the superposition of external forces (which is just $-\bm{F}_\mathrm{SI}$, because $\bm{F}_\mathrm{A}=\bm{0}$), the superposition of both surface tensions at the contact line and the superposition of surface tension gradients on the fluid:
\begin{equation}
  \label{eq:superposition}
  P_\mathrm{A+SI}=\int_{\mathcal{I}} \dd S \, (\bm{v}_\mathrm{A}+\bm{v}_\mathrm{SI})\cdot \boldsymbol{\nabla}_\mathrm{s}(\gamma_\mathrm{A} +\gamma_\mathrm{SI}) + \int_\ell \dd s \, (\gamma_\mathrm{A}+\gamma_\mathrm{SI}) \hat{\vect{n}}\cdot (\bm{V}_\mathrm{A}+\bm{V}_\mathrm{SI})  -\bm{F}_\mathrm{SI} \cdot (\bm{V}_\mathrm{A} +\bm{V}_\mathrm{SI})\,.
\end{equation}
 Two terms immediately cancel out because of the divergence theorem:
\begin{equation}
\int_{\mathcal{I}} \dd S \, \bm{v}_\mathrm{SI} \cdot \boldsymbol{\nabla}_\mathrm{s}\gamma_\mathrm{SI} + \int_\ell \dd s \, \gamma_\mathrm{SI} \hat{\vect{n}}\cdot \bm{V}_\mathrm{SI} = - \int_{\mathcal{I}} \dd S \, \gamma_\mathrm{SI} \boldsymbol{\nabla}_\mathrm{s}\cdot \bm{v}_\mathrm{SI} =0 \, .
\end{equation}
We now apply the Lorentz reciprocal theorem \citep{masoud2019} to the two problems:
\begin{equation}
  \int_{\mathcal{I}} \dd S \, \bm{v}_\mathrm{SI} \cdot \boldsymbol{\sigma}_\mathrm{A} \cdot \hat{\vect{e}}_z+ \int_{\mathcal{S}} \dd S\, \bm{v}_\mathrm{SI} \cdot \boldsymbol{\sigma}_\mathrm{A}(-\hat{\vect{n}})=  \int_{\mathcal{I}} \dd S \, \bm{v}_\mathrm{A} \cdot \boldsymbol{\sigma}_\mathrm{SI} \cdot \hat{\vect{e}}_z+ \int_{\mathcal{S}} \dd S\, \bm{v}_\mathrm{A} \cdot \boldsymbol{\sigma}_\mathrm{SI}(-\hat{\vect{n}})\,.
\end{equation}
By taking into account the force balance at the surface and at both bodies, we obtain
\begin{equation}
  \label{eq:lrt}
  \int_{\mathcal{I}} \dd S \, \bm{v}_\mathrm{SI} \cdot \boldsymbol{\nabla}_\mathrm{s}\gamma_\mathrm{A}+
  \int_\ell \dd s \, \gamma_\mathrm{A} \hat{\vect{n}}\cdot \bm{V}_\mathrm{SI} =
  \int_{\mathcal{I}} \dd S \, \bm{v}_\mathrm{A} \cdot \boldsymbol{\nabla}_\mathrm{s}\gamma_\mathrm{SI}+
  \int_\ell \dd s \, \gamma_\mathrm{SI} \hat{\vect{n}}\cdot \bm{V}_\mathrm{A} - \bm{F}_\mathrm{SI} \cdot \bm{V}_\mathrm{A}\,.
\end{equation}
At the same time, the divergence theorem applied to $\gamma_\mathrm{A} \bm{v}_\mathrm{SI}$ leads to
\begin{equation}
  \label{eq:divergencetheorem}
  \int_{\mathcal{I}} \dd S \, (\bm{v}_\mathrm{SI} \cdot \boldsymbol{\nabla}_\mathrm{s}\gamma_\mathrm{A} +\gamma_\mathrm{A} \boldsymbol{\nabla}_\mathrm{s} \cdot \bm{v}_\mathrm{SI}) 
  +
  \int_\ell \dd s \, \gamma_\mathrm{A} \hat{\vect{n}}\cdot \bm{V}_\mathrm{SI} = 0\,.
\end{equation}
Because we have chosen $\boldsymbol{\nabla}_\mathrm{s} \cdot \bm{v}_\mathrm{SI}=0$, Eq.~\eqref{eq:divergencetheorem} implies that both sides of Eq.~\eqref{eq:lrt} are zero.
The total contribution of all mixed terms in Eq.~\eqref{eq:superposition} therefore vanishes and the dissipation of the superposition flow
\begin{equation}
\label{eq:superposition2}
P_\mathrm{A+SI}=\int_{\mathcal{I}} \dd S \, \bm{v}_\mathrm{A}\cdot \boldsymbol{\nabla}_\mathrm{s}\gamma_\mathrm{A}  + \int_\ell \dd s \, \gamma_\mathrm{A} \hat{\vect{n}}\cdot \bm{V}_\mathrm{A}  
-\bm{F}_\mathrm{SI} \cdot \bm{V}_\mathrm{SI} = P_\mathrm{A}+P_\mathrm{SI}
\end{equation}
can be written as the sum of the dissipations in each of the problems.
This additivity is the second condition for deriving a minimum dissipation theorem for the active swimmer.

The derivation of the minimum dissipation theorem follows in the same way as for surface-driven swimmers \citep{Nasouri.Golestanian2021,DaddiMoussaIder.Vilfan2023}. We apply the inequality \eqref{eq:passmdt}  to the superposition 
\begin{equation}
\label{eq:ineq1}
P_\mathrm{A+SI} \ge (\bm{V}_\mathrm{A} +\bm{V}_\mathrm{SI}) \cdot \bm{R}_\mathrm{FS} \cdot (\bm{V}_\mathrm{A} +\bm{V}_\mathrm{SI})\,.
\end{equation}
This inequality holds for any values of the velocities, but we know that the equality is fulfilled if and only if the superposition represents exactly the flow around the passive body with free surface (FS problem).  This condition gives the following set of equations for the superposition
$\bm{V}_\mathrm{A}+\bm{V}_\mathrm{SI}=\bm{V}_\mathrm{FS}$ and $\bm{F}_\mathrm{SI}=
\bm{F}_\mathrm{FS}$. In the absence of translational-rotational coupling, the torque balance is satisfied automatically. At the same time, the drag forces in the FS and SI problem are related to their velocities through through the drag coefficients $\bm{R}_\mathrm{FS}$ and $\bm{R}_\mathrm{SI}$ as  $\bm{F}_\mathrm{FS}=- \bm{R}_\mathrm{FS}\cdot\bm{V}_\mathrm{FS}$ and  $\bm{F}_\mathrm{SI}=-\bm{R}_\mathrm{SI}\cdot\bm{V}_\mathrm{SI}$, which closes the equation system. Its solution is
\begin{subequations} \label{vFS_vSI}
	\begin{eqnarray}
		\label{vFS}
		\bm{V}_\mathrm{FS}&=&\left(\bm{I} -\bm{R}_\mathrm{SI}^{-1} \cdot\bm{R}_\mathrm{FS}\right)^{-1}\cdot\bm{V}_\mathrm{A},\\
		\label{vSI}
		\bm{V}_\mathrm{SI}&=&\left(\bm{R}_\mathrm{FS}^{-1} \cdot\bm{R}_\mathrm{SI}-\bm{I}\right )^{-1}\cdot\bm{V}_\mathrm{A}\;.
	\end{eqnarray}
\end{subequations}
By inserting these velocities into the inequality \eqref{eq:ineq1} we obtain the minimum dissipation theorem
\begin{equation}
  \label{eq:mdt-marangoni}
  P_\mathrm{A} \geq \bm{V}_\mathrm{A}\cdot\left(\bm{R}_\mathrm{FS}^{-1} -\bm{R}_\mathrm{SI}^{-1}\right )^{-1}\cdot\bm{V}_\mathrm{A} 
\end{equation}
which allows us to express a lower bound on the hydrodynamic dissipation by a Marangoni surfer with two drag coefficients for the horizontal motion of a body with the same shape in a fluid without surface tension gradients $\bm{R}_\mathrm{FS}$ and with surface incompressibility $\bm{R}_\mathrm{SI}$. Although we wrote the expressions for translational motion alone, it is straightforward to expand it to rotational motion, too. In this case, the generalised velocities become 3-component vectors with 2 translational and 1 rotational component and $\bm{R}_\mathrm{FS}$ and $\bm{R}_\mathrm{SI}$ become the corresponding grand resistance matrices. 

From the superposition, we also obtain the  flow field of the optimal Marangoni surfer as $\bm{v}_\mathrm{A}=\bm{v}_\mathrm{FS}-\bm{v}_\mathrm{SI}$ and its relative surface tension $\gamma_\mathrm{A}=-\gamma_\mathrm{SI}$. Here both passive solutions are evaluated for bodies moving with the velocities as determined by Eqs.~\eqref{vFS_vSI}. Along with the bound on dissipation, the minimum dissipation theorem also provides us with the solution for the surface tension of the active Marangoni surfer that will minimise the hydrodynamic dissipation.

\subsection{Minimum dissipation for a circular disk-shaped Marangoni surfer}

In the following, we employ the minimum dissipation theorem~\eqref{eq:mdt-marangoni} to a thin Marangoni surfer with the shape of a circular disk. 
For this geometry, both drag coefficients are known from the literature. For an incompressible surface, it can be found as a limiting case in the solutions of 
\citet{Hughes.White1981} and \citet{Stone.Ajdari1998}:
\begin{equation}
  R_\mathrm{SI} = 8\mu a \, . \label{eq:drag_SI}
\end{equation}
The solution with a free planar surface is mathematically equivalent to the edgewise motion of a thin disk in bulk fluid \citep{happel1983}, with half the drag coefficient:
\begin{equation}
  R_\mathrm{FS} = \frac{16}{3} \, \mu a \, . \label{eq:drag_FS}
\end{equation}
      
With these drag coefficients,
the minimum dissipation theorem gives us a lower bound on the hydrodynamic dissipation by a disk-shaped Marangoni surfer as
\begin{equation}
  P_\mathrm{A} \ge 16\mu a V_\mathrm{A}^2\,.
\end{equation}
The Lighthill efficiency \citep{lighthill1952}, defined as the ratio between the dissipation by a passive swimmer steadily towed by an external force and the active swimmer, has the upper limit
\begin{equation}
  \label{eq:lighthill}
  \eta_\mathrm{L}=\frac{R_\mathrm{FS}V_\mathrm{A}^2 }{P_\mathrm{A}} \le \frac 1 3
\end{equation}
Unlike in surface-slip driven microswimmers \citep{leshansky2007}, the Lighthill efficiency is always bounded by~1. A hydrodynamic efficiency of $1/3$ places the Marangoni surfers to the top of realistic swimmer designs. Although an ideal spherical surface driven swimmer has an efficiency limit of $\eta_\mathrm{L}\le 1/2$ \citep{michelin2010,guo2021}, which is even higher for elongated swimmers \citep{Nasouri.Golestanian2021}, such high efficiency requires a frictionless mechanism of slip generation, which is difficult to realise.

The full solution of the optimal disk-shaped surfer that includes surface tension, fluid velocity and forces can be obtained as a superposition of these fields in the two passive problems. For this purpose, we derive both solutions using a uniform notation in sections \ref{sec:3} and \ref{sec:4}.

\subsection{Approximate solution for a surfer of finite depth}

If the surfer has a finite depth, it is possible to derive a lower bound on the dissipation in a perturbative way. Specifically, we treat a surfer with the shape of an oblate spheroid with major semi-axis $a$ and minor semi-axis $\varepsilon a$. The spheroid is half-submerged, such that its surface $\mathcal S$ is given by the equation $z=-\varepsilon \sqrt{a^2-\rho^2}$ in cylindrical coordinates.

Perturbative solutions for both drag coefficients, $R_\mathrm{FS}$ and $R_\mathrm{SI}$, were obtained by \citet{Stone.Masoud2015}. In the case of a force-free surface, the drag coefficient of the half-submerged spheroid is $1/2$ the coefficient for edgewise translation in bulk fluid, which is known exactly \citep{happel1983}. For thin spheroids, its expansion reads
\begin{equation}
	R_\mathrm{FS} = \frac{16}{3} \mu a \left( 1 + \frac{8}{3\pi} \, \varepsilon  \right) + \mathcal{O} \left( \varepsilon^2 \right)\,.
\end{equation}
The solution in the presence of an incompressible interface was formulated using the Lorentz reciprocal theorem, with the solution for a thin disk serving as the auxiliary problem \citep{Stone.Masoud2015}. An analytical expression can be obtained by integrating Eq.~(3.6) from \citep{Stone.Masoud2015} in the limit $Bq=0$: 
\begin{equation}
	R_\mathrm{SI} = 8\mu a \left( 1 + \frac{4\left( 1-\ln 2 \right)}{\pi} \, \varepsilon \right) + \mathcal{O} \left( \varepsilon^2 \right)\,.
\end{equation}
The minimum dissipation follows from (\ref{eq:mdt-marangoni}) as
\begin{equation}
  P_\mathrm{A} \geq (R_\mathrm{FS}^{-1}-R_\mathrm{SI}^{-1})^{-1} = 16\mu a V_\mathrm{A}^2 \left( 1 + \frac{8 \ln 2}{\pi} \,  \varepsilon \right)
	+ \mathcal{O} \left( \varepsilon^2 \right)
\end{equation}
and the upper bound on Lighthill efficiency is
\begin{equation}
  \eta_\mathrm{L} \leq 1-\frac{R_\mathrm{FS}}{R_\mathrm{SI}}= \frac{1}{3} - \frac{8}{9\pi} \left( 3\ln 2-1 \right) \varepsilon + \mathcal{O} \left( \varepsilon^2 \right)\, .
\end{equation}
The finite depth of the swimmer not only increases the dissipation, but also reduces its Lighthill efficiency. Similar asymptotic solutions are also possible for other swimmer shapes, for example for a partially submerged sphere, with the submerged surface $\mathcal S$ taking the shape of a spherical cap \citep{Stone.Masoud2015}. 

\section{Thin circular disk translating in a fluid with surface incompressibility}
\label{sec:3}

In this section, we derive the hydrodynamic field resulting from the translational movement of a passive circular disk positioned at an interface with surface incompressibility. 
Theoretical studies of the flow past a disk embedded in an incompressible viscous surface, representing a model of a lipid bilayer, date back to \citet{saffman1975brownian}. 
\citet{saffman1976brownian} examined the limit ${Bq} \gg 1$ where the Boussinesq number is defined as ${Bq} = \mu_\mathrm{s} / \left( a\mu\right)$ and $\mu_\mathrm{s}$ represents the surface viscosity. In this limit, the membrane viscosity dominates over the viscosity of the surrounding fluid.
\citet{Hughes.White1981} solved the model for arbitrary viscosities, providing analytical expressions for the viscous flow field and resistance coefficients through the use of dual integral equations.
The problem with a finite subphase depth was subsequently solved by \cite{Stone.Ajdari1998}, partly through a numerical approach.
For an infinite depth and vanishing surface viscosity (${Bq} \to 0$), both solutions give a translational drag coefficient $8\mu a$, which is be 50\% larger than the drag coefficient $16/3\mu a$ in a fluid with free surface. More recently, \cite{yariv2023motion} revisited the translational motion of a disk embedded in a nearly inviscid Langmuir film, i.e., in the limit ${Bq} \ll 1$.

Because the solutions in \cite{Hughes.White1981} and \cite{Stone.Ajdari1998} are derived for arbitrary $Bq$ and have a more complex form, we are rederiving the solution without surface viscosity in the following in order to obtain expressions for the velocity and force fields that are more suitable for practical usage. In the following we drop the `SI' subscript as all quantities involved belong to the problem with surface incompressibility.

\subsection{Green's functions in Fourier space}

We begin by solving the flow equations~\eqref{eq:fluid-motion-UND-incompressibility} for a given distribution of surface forces.
The impermeability condition requires zero normal velocity at the interface, i.e., $v_z = 0$ at $z=0$. The force balance at the interface in the in-plane direction states:
\begin{equation}
	\left. \mu \partial_z \vect{v}_\parallel  \right|_{z=0}
	=
\begin{cases}
        \vect{f}_\parallel &\text{for}\qquad \rho<a \\ \boldsymbol{\nabla}_\mathrm{s} \gamma &\text{for}\qquad \rho>a \, ,
      \end{cases}
      \label{eq:condition-force-surf-tension-at-inter}
      \end{equation}
wherein $\vect{f}_\parallel$ is an unknown in-plane surface force density acting on the fluid from the surface of the disk.
Finally, we require the in-plane divergence at the interface to be zero:
\begin{equation}
	\left. \boldsymbol{\nabla}_\mathrm{s} \cdot \vect{v}_\parallel \right|_{z=0} = 0 \, . \label{eq:tan-grad-v-at-inter}
\end{equation}
Together with the incompressibility equation~\eqref{eq:incompressibility}, it follows that 
\begin{equation}
	\left. \partial_z v_z \right|_{z=0} = 0 \, . \label{eq:condition-div-vz}
\end{equation}

In order to have a more uniform boundary condition at the surface, we arbitrarily extend the surface tension difference $\gamma$ to the region underneath the disk ($\rho<a$) while requiring continuity at $\rho=a$. At the same time, we introduce a transformed force density
\begin{equation}
  \label{eq:fredefined}
  \vect{F}_\parallel=\begin{cases} 
    \vect{f}_\parallel-\boldsymbol{\nabla}_\mathrm{s} \gamma & \text{for}\quad \rho<a\\
    \bm{0} & \text{for}\quad \rho>a
  \end{cases}
\end{equation}
With this redefinition, Eq.~\eqref{eq:condition-force-surf-tension-at-inter} obtains the unified form
\begin{equation}
  \left. \mu \partial_z \vect{v}_\parallel \right|_{z=0}
  =
  \vect{F}_\parallel + \boldsymbol{\nabla}_\mathrm{s} \gamma\;.
  \label{eq:condition-force-surf-tension-at-inter-a}
\end{equation}
Equation \eqref{eq:tan-grad-v-at-inter} is also trivially satisfied under the disk and therefore in the whole plane.

To determine the solution for the flow velocity and pressure fields, we use a 2D Fourier transform along the $x$ and $y$ directions.
We define the forward 2D Fourier transform of a given function $f(\boldsymbol{\rho}, z)$ as 
\begin{equation}
	\widetilde{f}(\bm{k},z) = \mathcal{F}\left\{ f(\boldsymbol{\rho},z) \right\} = \int_{\mathbb{R}^2} \mathrm{d}^2 \boldsymbol{\rho} \, f(\boldsymbol{\rho},z) \, e^{-i \bm{k} \cdot \boldsymbol{\rho}} \, ,
\end{equation}
where $\bm{k}$ represents the wavevector. The $z$-dependence is left unchanged by the transformation. 

Similarly, we define the inverse 2D Fourier transform as:
\begin{equation}
	f(\boldsymbol{\rho},z) = \mathcal{F}^{-1}\left\{ \widetilde{f}(\bm{k},z) \right\} = \frac{1}{\left(2\pi\right)^2} \int_{\mathbb{R}^2} \mathrm{d}^2 \bm{k} \, \widetilde{f}(\bm{k}, z) \, e^{i \bm{k} \cdot \boldsymbol{\rho}} \, .
\end{equation}
In these equations, $\boldsymbol{\rho} = (x, y)$ denotes the position vector along the interface.
We also introduce the wavenumber $ k = |\vect{k}|$, which represents the magnitude of the wavevector, and define the unit vector $\hat{\vect{k}} = \vect{k}/k$.

\subsubsection{Normal velocity}

By forming the divergence of Eq.~(\ref{eq:fluid-motion-UND-incompressibility}a) and applying the incompressibility condition, it follows that the pressure field is harmonic, $\nabla^2 p = 0$ \citep{happel1983}.
Thus, the velocity field adheres to the biharmonic equation $\boldsymbol{\nabla}^4 \vect{v} = 0$, which can be represented in Fourier space as 
\begin{equation}
	\left( \partial_z^2 - k^2 \right)^2 \widetilde{\vect{v}} = \vect{0} \, . \label{eq:biharmonic-Fourier}
\end{equation}
This equation represents a homogeneous fourth-order linear differential equation for $\widetilde{\vect{v}}$ and is solved by
\begin{equation}
	\widetilde{\vect{v}} = \left( \boldsymbol{\alpha}_1 + z \boldsymbol{\alpha}_2 \right) e^{kz} \, ,
\end{equation}
with the unknown wavenumber-dependent vector functions $\boldsymbol{\alpha}_1$ and~$\boldsymbol{\alpha}_2$ that will be determined from the boundary conditions.
We have retained only the solution with the decaying exponential to meet the regularity condition of finite velocity as $z \to -\infty$.

Given that both the normal velocity at the interface, $\widetilde{v}_z$, and its derivative with respect to~$z$, $\partial_z \widetilde{v}_z$, are zero at the interface, c.f.\ Eq.~\eqref{eq:condition-div-vz}, we can conclude that $\widetilde{v}_z = 0$ throughout. Therefore, the velocity field only has an in-plane component, denoted henceforth as $\widetilde{\vect{v}}_\parallel$.

\subsubsection{In-plane velocity}

In the following we determine the solution for the in-plane component $\widetilde{\vect{v}}_\parallel$ of the velocity.
In Fourier space, the flow equations \eqref{eq:fluid-motion-UND-incompressibility} are projected as
\begin{subequations} \label{eq:fluid-motion-UND-incompressibility-Fourier}
	\begin{eqnarray}
		- i \vect{k} \widetilde{p} + \mu \left( \partial_z^2 - k^2 \right) \widetilde{\vect{v}}_\parallel &=& \bm{0} \, , \label{eq:fluid-motion-Fourier} \\
		\vect{k} \cdot \widetilde{\vect{v}}_\parallel &=& 0  \label{eq:incompressibility-Fourier}
	\end{eqnarray}
\end{subequations}
and the force balance at the interface \eqref{eq:condition-force-surf-tension-at-inter-a} as
\begin{equation}
	\left. \mu \partial_z \widetilde{\vect{v}}_\parallel \right|_{z=0} = \widetilde{\vect{F}}_\parallel + i \vect{k} \widetilde{\gamma} \, . \label{eq:condition-force-surf-tension-at-inter-Fourier}
\end{equation}
By multiplying both sides of Eq.~\eqref{eq:fluid-motion-Fourier} with $\hat{\vect{k}}$ and using Eq.~\eqref{eq:incompressibility-Fourier}, we see that the pressure vanishes, $\widetilde{p} = 0$.
Consequently, \eqref{eq:fluid-motion-Fourier} simplifies to
\begin{equation}
	\left( \partial_z^2 - k^2 \right) \widetilde{\vect{v}}_\parallel = \vect{0} \, , 
\end{equation}
which is a homogeneous second-order linear differential equation. Its solution that does not diverge for $z\to -\infty$ 
is given by
\begin{equation}
	\widetilde{\vect{v}}_\parallel = \boldsymbol{\alpha}_3 e^{kz} \, , \label{eq:v_para}
\end{equation}
where~$\boldsymbol{\alpha}_3$ is a wavenumber-dependent function that can be determined from the boundary condition \eqref{eq:condition-force-surf-tension-at-inter-Fourier} as
\begin{equation}
	\boldsymbol{\alpha}_3 = \frac{1}{\mu k} \left( \widetilde{\vect{F}}_\parallel + i \vect{k} \widetilde{\gamma} \right) .
	\label{eq:alpha3}
\end{equation}

The condition of vanishing in-plane divergence at the interface \eqref{eq:tan-grad-v-at-inter} reads $\vect{k} \cdot \boldsymbol{\alpha}_3 = 0$ in Fourier space. It allows us to express the surface tension with the force density as
\begin{equation}
	\widetilde{\gamma} = \frac{i \vect{k} \cdot \widetilde{\vect{F}}_\parallel}{k^2} \, . 
	\label{eq:surf-tension-fourier}
\end{equation}

Finally, by substituting Eq.~\eqref{eq:surf-tension-fourier} into Eq.~\eqref{eq:alpha3} and Eq.~\eqref{eq:v_para}, the solution for the velocity is 
\begin{equation}
	\widetilde{\vect{v}}_\parallel = \frac{e^{kz}}{\mu k} \left( \vect{I} - \hat{\vect{k}} \,\hat{\vect{k}} \right)\cdot
	\widetilde{\vect{F}}_\parallel \, .
	\label{eq:v_para-final}
\end{equation}
The velocity therefore only has a component that is parallel to the surface and transverse to the wavevector. A more general solution allowing for a finite thickness of the fluid layer and surface viscosity can be found, for instance, in \cite{PhysRevX.11.031065}.

\subsection{Solution for the circular disk}

To solve the flow around a circular disk, we use a cylindrical coordinate system and express the unit wavevector $\hat{\vect{k}}$ as $\hat{k}_x = \cos\phi$ and $\hat{k}_y = \sin\phi$.
In addition, we define the unit vector~$\hat{\vect{t}}$ along the direction perpendicular to the wavevector such that $\hat{t}_x = \sin\phi$ and $\hat{t}_y = -\cos\phi$.
Accordingly, the Fourier transform of the surface force density can be expressed in the unit vector basis composed of $\hat{\vect{k}}$ and $\hat{\vect{t}}$ as
\begin{equation}
	\widetilde{\vect{F}}_\parallel = \widetilde{F}_l \, \hat{\vect{k}} + \widetilde{F}_t \, \hat{\vect{t}} \, , 
	\label{eq:F}
\end{equation}
wherein $\widetilde{F}_l$ and~$\widetilde{F}_t$ stand for the longitudinal and transverse components of the surface force density, respectively \citep{bickel2007hindered}.
Consequently, we can express the in-plane component of the flow velocity in Fourier space by referring to Eq.~\eqref{eq:v_para-final} as
\begin{equation}
	\widetilde{\vect{v}}_\parallel = \frac{e^{kz}}{\mu k} \, \widetilde{F}_t \, \hat{\vect{t}} \, .
	\label{eq:v_para-final-disk}
\end{equation}

Without loss of generality we assume that the disk moves along the positive $x$ direction.
We anticipate the angular structure of the Fourier-transformed surface force density to take the following form:
\begin{subequations} \label{eq:Ansatz_Kraefte_Fourier}
	\begin{eqnarray}
		\widetilde{F}_l (k,\phi) &=& \widetilde{M}(k) \cos\phi \, , \\
		\widetilde{F}_t (k,\phi) &=& \widetilde{N}(k) \sin\phi \, , 
	\end{eqnarray}
\end{subequations}
where $\widetilde{M} (k)$ and $\widetilde{N} (k)$ are unknown wavenumber-dependent functions to be determined from the underlying boundary conditions.
We made this specific selection because the inverse Fourier transform will result in an angular structure akin to the one required for matching in the boundary conditions imposed at the surface of the disk.
Inverse Fourier transform of Eqs.~\eqref{eq:Ansatz_Kraefte_Fourier} yields the radial and tangential components of the surface force density as
\begin{subequations} \label{eq:forces}
	\begin{eqnarray}
		F_\rho &=& \frac{\cos\theta}{4\pi} \int_0^\infty k \, \mathrm{d} k
		\left[ \left( \widetilde{M}(k)+\widetilde{N}(k) \right) \mathrm{J}_0(\rho k) - \left( \widetilde{M}(k)-\widetilde{N} (k) \right) \mathrm{J}_2(\rho k) \right]  \, , \qquad \\
		F_\theta &=& -\frac{\sin\theta}{4\pi} \int_0^\infty k \, \mathrm{d} k
		\left[ \left( \widetilde{M}(k)+\widetilde{N} (k)\right) \mathrm{J}_0(\rho k) + \left( \widetilde{M}(k)-\widetilde{N}(k) \right) \mathrm{J}_2(\rho k) \right] \, .
	\end{eqnarray}
\end{subequations}
Likewise, the velocities induced by this force distribution transform to real space as
\begin{subequations} \label{eq:flow-field-inter}
	\begin{eqnarray}
		v_\rho &=&  \frac{\cos\theta}{4\pi\mu} \int_0^\infty \mathrm{d} k \, \widetilde{N} (k) \left( \mathrm{J}_0(\rho k) + \mathrm{J}_2(\rho k) \right) e^{kz} \, , \\
		v_\theta &=& -\frac{\sin\theta}{4\pi\mu} \int_0^\infty \mathrm{d} k \, \widetilde{N} (k) \left( \mathrm{J}_0(\rho k) - \mathrm{J}_2(\rho k) \right) e^{kz} \, .
	\end{eqnarray}
\end{subequations}

The surface tension, given by  Eq.~\eqref{eq:surf-tension-fourier}, can be expressed in terms of the longitudinal component of the surface force density as
\begin{equation}
	\widetilde{\gamma} (k, \phi) = \frac{i}{k} \, \widetilde{F}_l (k, \phi) = \frac{i}{k} \, \widetilde{M} (k) \cos\phi \, ,
\end{equation}
which can be transformed back to real space through an inverse transformation as
\begin{equation}
  \gamma = -\frac{\cos \theta}{2\pi} \int_0^\infty \mathrm{d}k \, \widetilde{M} (k) \, \mathrm{J}_1(\rho k) \, .
\label{eq:gammainversetrans}  
\end{equation}

\subsubsection{Force density}

After representing the flow velocity as integrals involving wavenumber-dependent functions, the next step involves the formulation of dual integral equations. 
The integral equations for the inner domain are obtained from the no-slip boundary conditions at the surface of the disk,  $v_\rho = V \cos\theta$ and $v_\theta = -V\sin\theta$ for $z=0$ and $\rho \in [0, a]$:
\begin{equation} \label{eq:inner-problem}
		\int_0^\infty \mathrm{d}k \, \widetilde{N}(k) \, \mathrm{J}_0(\rho k) = 4\pi\mu V \, ,  \qquad \int_0^\infty \mathrm{d}k \, \widetilde{N}(k) \, \mathrm{J}_2(\rho k) = 0 \, .
\end{equation}
The integral equations for the outer problem are obtained by imposing a vanishing surface force density \eqref{eq:forces} at the surface outside the disk for $z=0$ and $\rho > a$:
\begin{equation} \label{eq:outer-problem}
  \int_0^\infty k \, \mathrm{d}k \left( \widetilde{M}(k) +  \widetilde{N}(k) \right) \mathrm{J}_{0}(\rho k) = 0 \, ,
  \quad
    \int_0^\infty k \, \mathrm{d}k \left( \widetilde{M}(k) - \widetilde{N}(k) \right) \mathrm{J}_{2}(\rho k) = 0 \, .
\end{equation}

Equations~\eqref{eq:inner-problem} and~\eqref{eq:outer-problem} form a system of dual integral equations on the inner and outer domain boundaries.
The solution can be obtained using established techniques as outlined in the works of \cite{sneddon60, sneddon66} and \cite{copson1947problem, copson61}.
The basic idea involves seeking a set of solutions that satisfy the equations for the outer problems \eqref{eq:outer-problem}. Typically, the solution for the unknown wavenumber-dependent functions is explored through definite integrals, often resulting in vanishing values in the outer domain when $\rho > a$. For detailed solution approaches, one can refer to Chapter IV of \cite{sneddon66} for dual integral equations and Chapter VIII for their applications in electrostatics.
When incorporating these solution forms into the equations for the inner problem \eqref{eq:inner-problem}, one encounters a system of Fredholm integral equations, the solution of which is not always straightforward to obtain. 
In our case, we employ the power series method to derive the solution, whereby the unknown function is expanded in terms of unspecified coefficients. 
Subsequently, we evaluate the resulting integrals analytically. Upon determining the coefficients through comparison with the known right-hand side, the unknown function is then discerned from its series expansion. For further details, refer to Appendix of \cite{daddi2020dynamics} where a similar approach was employed.
However, in similar scenarios, it is anticipated that the solution may involve combinations of trigonometric and Bessel functions. Therefore, employing a guessed solution can often be advantageous in the given context. For instance, this approach has been applied to the flow generated by different types of force or source singularities near circular interfaces \citep{daddi2021steady, daddi2022diffusiophoretic, PhysRevResearch.5.033030}. In our case, the dual integral equations are solved by:
\begin{subequations}
	\begin{eqnarray}
		\widetilde{M}(k) &=& \frac{16\mu V}{k} \, \mathrm{J}_1(ka) \, , \\
		\widetilde{N}(k) &=& \frac{8\mu V}{k} \, \sin (ka) \, .
	\end{eqnarray}
\end{subequations}

The solutions we derived can now be inserted into the integrals that stem from the inverse Fourier transforms, \eqref{eq:forces}, \eqref{eq:flow-field-inter} and \eqref{eq:gammainversetrans}. 
The relative surface tension follows from Eq.~\eqref{eq:gammainversetrans}
\begin{equation} 
  \gamma = -\frac{4\mu V}{\pi} \, \cos\theta \, \times
  \begin{cases}
    \frac{\rho}{a} & \text{if} \quad \rho < a \, , \\
    \frac{a}{\rho} & \text{if} \quad \rho > a \, .
  \end{cases}
  \label{eq:gamma}
\end{equation}
The gradient of the surface tension around the disk is therefore proportional to the inverse square of the distance. 
The expression for the interior, of course, only represents the surface tension arbitrarily extended across the disk with no physical meaning. By inserting it into Eq.~\eqref{eq:fredefined}, it gives the physical force density exerted on the fluid by the disk
\begin{equation}
  \vect{f}_\parallel= \vect{F}_\parallel -\frac{4}{\pi} \frac{\mu V}{a} \, \hat{\vect{e}}_x \, .
\end{equation}

After carrying out the integrals given in  Eqs.~\eqref{eq:forces}, we can express the resulting force density at the surface of the disk in Cartesian coordinates as:
\begin{equation}
	\vect{f}_\parallel = \frac{4\mu V}{\pi a} \left\{  \left[ A(\rho) + B(\rho)\cos (2\theta) \right] \hat{\vect{e}}_x
	+ B(\rho) \sin (2\theta) \, \hat{\vect{e}}_y \right\} \, , \label{eq:Kraft}
\end{equation}
where we have defined the dimensionless radial functions
\begin{subequations}
	\begin{eqnarray}
		A(\rho) &=&  \frac{a}{2 \sqrt{a^2-\rho^2}}, \\
		B(\rho) &=& - \left( \frac{a}{\rho} \right)^2 
		\left( \frac{2a^2-\rho^2}{2a \sqrt{a^2-\rho^2}} - 1 \right) .
	\end{eqnarray}
\end{subequations}
It can be seen that $A$ is a monotonically increasing function of~$\rho$ that starts with the value of $1/2$ at the origin, while $B$ is a monotonically decreasing function of~$\rho$, starting from zero at the origin.
It is evident that both $A$ and $-B$ exhibit asymptotic divergence, scaling to leading order as $1/\sqrt{a-\rho}$ as $\rho$ approaches $a$.
The maximum magnitude of the surface force density is attained at $A - B$ when $\theta \in \{\pi/2, 3\pi/2\}$, while the minimum magnitude occurs at $A + B$ when $\theta \in \{0, \pi\}$.

The drag force on the disk can be calculated as the sum of the shear forces at the bottom surface and the surface tension on the perimeter,
\begin{equation}
  \bm{F}_\mathrm{SI}=-\int_{\mathcal{S}} \dd S \, \vect{f}_\parallel - \int_\ell  \dd s \, \gamma \hat{\vect{n}}\;.
\end{equation}
The contribution of surface tension \eqref{eq:gamma} amounts to $-4\mu a V$. The integral of the surface force density \eqref{eq:Kraft}, to which only the term $A(\rho)$ contributes, equally gives $-4\mu a V$. Together, the drag coefficient of a thin circular disk in a fluid with surface incompressibility \eqref{eq:drag_SI} is obtained as
\begin{equation}
	R_\mathrm{SI} = 8\mu a 
\end{equation}
and consists of equal contributions by the shear stress and by the surface tension. The drag coefficient agrees with the limiting cases from the expressions by 
\citet{Hughes.White1981} and \citet{Stone.Ajdari1998}. 

\begin{figure}
  \begin{center}
    \includegraphics[width=0.65\textwidth]{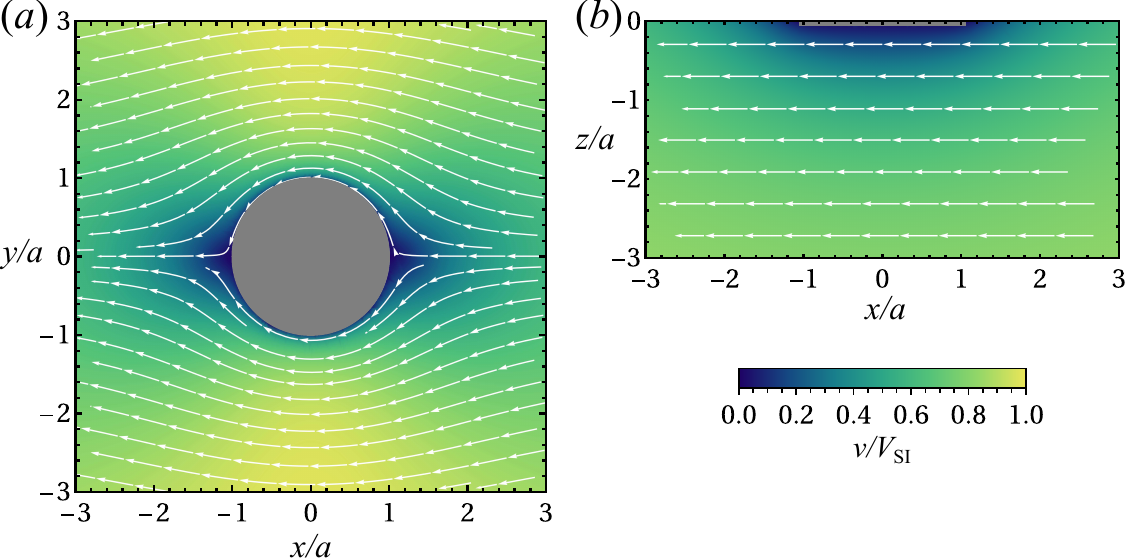}
  \end{center}
  \caption{\label{fig:3}The flow field of a circular disk embedded in an incompressible interface (SI problem), shown in the co-moving frame. The top view at the surface is shown in panel~(\textit{a}) and side view at $y=0$ in panel~(\textit{b}).}
\end{figure}

\subsection{Flow field}

To derive a closed analytical expression for the flow field, determined by Eqs.~\eqref{eq:flow-field-inter}, we define the sequence of infinite integrals
\begin{equation}
	C_n  = \int_0^\infty \mathrm{d} k \, \frac{\sin \left( ka\right)}{k} \, \mathrm{J}_n(\rho k) \, e^{kz}  \, ,
	\label{eq:C_n}
\end{equation}
for $z \le 0$ to ensure convergence.
It can be shown that $C_0 = \operatorname{arg} \left\{  Q \right\}$ and $C_2 = \operatorname{Im} \left\{ s Q \right\}$, where we have defined the abbreviation $Q = s + S$ with $s=\left( z+ia \right)/\rho$ and $S = \sqrt{1+s^2}$.
Here, `arg' represents the argument of a complex number, while `Im' signifies its imaginary part.
To obtain these results, it is sufficient to represent the sine function using Euler's notation and apply the Laplace transform to the Bessel function \citep{watson1922treatise}. 
For a detailed treatment of integrals with similar forms, see the Appendix in \cite{daddi2020axisymmetric}.
The radial and azimuthal components of the flow field can then be cast in a compact form as
\begin{subequations}
	\begin{eqnarray}
		\frac{v_\rho}{V} &=& \frac{2}{\pi} \left( C_0 + C_2 \right) \cos\theta \, , \\
		\frac{v_\theta}{V} &=& -\frac{2}{\pi} \left( C_0 - C_2 \right) \sin\theta \, .
	\end{eqnarray}
\end{subequations}

At the boundary, where $z=0$, a closed-form expression for $C_n$ can be derived for any arbitrary~$n \ge 0$ as
\begin{equation}
  C_n (z=0) = \begin{cases}
    \frac{1}{n} \left( \frac{\rho}{a + \sqrt{a^2-\rho^2}} \right)^n \sin \left( \frac{n\pi}{2} \right) & \text{if} \quad \rho < a \, , \\
    \frac{1}{n} \sin \left[ n \arcsin \left( \frac{a}{\rho} \right) \right] & \text{if} \quad \rho > a \, .
  \end{cases}
\end{equation}
In particular, for $\rho < a$, it can be noticed that $C_{2n}(z=0) = 0$ for $n \ge 1$. For $n=0$, it follows that $C_0(z=0) = \pi/2$ if $\rho < a$ and $C_0(z=0) = \arcsin \left( a/\rho \right)$ if $\rho > a$.

The flow field, transformed into the co-moving frame ($(\bm{v}-\hat{\vect{e}}_x V)/V$) is shown in Fig.~\ref{fig:3}.

\section{Thin circular disk translating in a fluid with force-free surface}
\label{sec:4}

In the following, we recapitulate the solution for the flow and the drag force on a thin circular disk, moving horizontally at a gas-liquid interface. Because the interface is assumed to be planar and is not affected by the flow, the solution is mathematically equivalent to the edgewise motion of a thin disk in bulk fluid, with half the drag coefficient. The latter has been solved by \citet{ray1936application} using Bessel functions. The flow around a thin disk can also be obtained from the solution for an ellipsoidal particle in the limit of zero thickness \citep{lamb1945hydrodynamics}. 
A general approach for analysing the arbitrary motion of a circular disk in a Stokes flow has later been formulated using dual integral equation approach \citep{tanzosh1994integral, Tanzosh.Stone1996}. 
The problem of a disk moving sideways between parallel walls has further been investigated by \citet{davis1991slow}, who determined the additional drag encountered by the disk by solving a pair of integral equations~\citep{davis1991slow}.
In order to ensure uniform notation needed for the superposition of both solutions, we give a brief summary of the of the derivation by \citet{Tanzosh.Stone1996}  using our notation in the following. We will omit the `FS' subscript as all quantities in this section are related to a disk in motion at a free interface.

We express the forces in Fourier space using Eq.~\eqref{eq:Ansatz_Kraefte_Fourier}. Since this problem has been thoroughly examined in the literature, we refrain from presenting the complete solution steps here. 
In the case of a free surface, our expectation is that the force density aligns along the direction of motion.
Unlike in the case of an incompressible surface, where $\widetilde{\vect{v}}_\parallel$ is found to be dependent solely on $\widetilde{F}_t$, see Eq.~\eqref{eq:v_para-final-disk}, the absence of surface tension in the in-plane force balance introduces a different dependence. In this scenario, the in-plane velocity is determined by both $\widetilde{F}_t$ and $\widetilde{F}_l$.
By anticipating that the force is aligned along the $x$ direction, one would posit $\widetilde{M}(k) = \widetilde{N}(k)$. It is worth noting that a systematic investigation without enforcing this condition leads to the same conclusion.
By requiring no-slip boundary conditions at the surface of the disk, the corresponding dual integral integrals for the inner domain for $z=0$ and $\rho \in [0,a]$ are
\begin{equation}
		\int_0^\infty \mathrm{d}k \, \widetilde{M}(k) \, \mathrm{J}_{0}(\rho k) = \frac{8}{3} \, \pi\mu V \, , \quad  
		\int_0^\infty \mathrm{d}k \, \widetilde{M}(k) \, \mathrm{J}_{2}(\rho k) = 0 \, ,
\end{equation}

Equations~\eqref{eq:outer-problem} on the outer domain remain the same as these are connected to the absence of force beyond the disk.
Then, the solution of the resulting dual integral equations is obtained as
\begin{equation}
	\widetilde{M}(k) = \widetilde{N}(k) = \frac{16 }{3} \frac{\mu V}{k} \, \sin (ka) \, .
\end{equation}

In contrast to the case of incompressible surface, the expression for the surface force density assumes a straightforward form and can be derived from Eq.~\eqref{eq:forces} as
\begin{equation}
	\vect{f}_\parallel = \frac{8\mu V}{3\pi \sqrt{a^2-\rho^2}} \, \hat{\vect{e}}_x \, .
\end{equation}
Integrating this force over the surface of the disk leads to the familiar expression for the drag coefficient \eqref{eq:drag_FS}
\begin{equation}
	R_\mathrm{FS} = \frac{16}{3} \, \mu a \, . 
\end{equation}

The hydrodynamic flow field is determined by carrying out the inverse Fourier transform, resulting in the components of the velocity field as follows:
\begin{subequations}
	\begin{eqnarray}
		\frac{v_\rho}{V} &=& \frac{2}{3\pi} \left[  3 C_0 + C_2 + \left( D_0 - D_2 \right) \frac{z}{\rho} \right] \cos\theta \, , \\
		\frac{v_\theta}{V} &=& -\frac{2}{3\pi} \left[  3 C_0 - C_2 + \left( D_0 + D_2 \right) \frac{z}{\rho} \right] \sin\theta , \\
		\frac{v_z}{V} &=& \frac{4}{3\pi} \frac{z}{\rho} \, D_1 \cos\theta \, .
	\end{eqnarray}
\end{subequations}
The sequence $C_n$ was previously defined in terms of an infinite integral by Eq.~\eqref{eq:C_n}.
Here, $D_n = \rho \partial_z C_n$, also a dimensionless function.
It can be shown that $D_0 = \operatorname{Im} \{ 1/S \}$, $D_1 = \operatorname{Im} \{s/ S\}$, and $D_2 = \operatorname{Im} \{ Q^2/ S\}$.
When $z=0$, a closed-form expression for $D_n$ can likewise be derived for any~$n \ge 0$ as
\begin{equation}
	D_n (z=0) = \begin{cases}
		\frac{\rho}{\sqrt{a^2-\rho^2}} \left( \frac{\rho}{a + \sqrt{a^2-\rho^2}} \right)^n \cos \left( \frac{n\pi}{2} \right) & \text{if} \quad \rho < a \, , \\
		\frac{\rho}{\sqrt{\rho^2-a^2}} \sin \left[ n \arcsin \left( \frac{a}{\rho} \right) \right]  & \text{if} \quad \rho > a \, .
	\end{cases}
\end{equation}
In particular, for $z=0$ and $\rho < a$, all odd terms vanish, $D_{2n+1} = 0$.

\begin{figure}
  \begin{center}
    \includegraphics[width=0.65\textwidth]{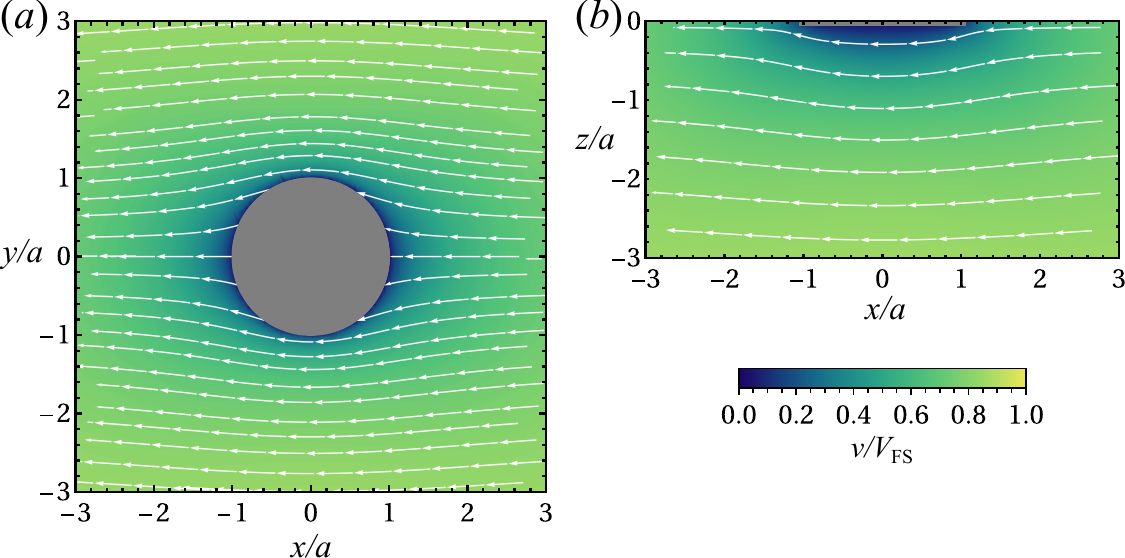}
  \end{center}
  \caption{\label{fig:4}The flow field of a circular disk embedded in a force-free interface (FS problem), shown in the co-moving frame in top view (\textit{a}) and side view (\textit{b}).}
\end{figure}

Figure \ref{fig:4} shows the resulting flow at the surface and in a vertical cross-section. Unlike with an interface with surface incompressibility, where the pressure is zero, the solution for a free surface includes a non-zero pressure field, which is related to the normal velocity as $p = 2\mu v_z/z$.

\section{The optimal propulsion of a disk-shaped Marangoni surfer}

\begin{figure}
  \begin{center}
    \includegraphics[width=\textwidth]{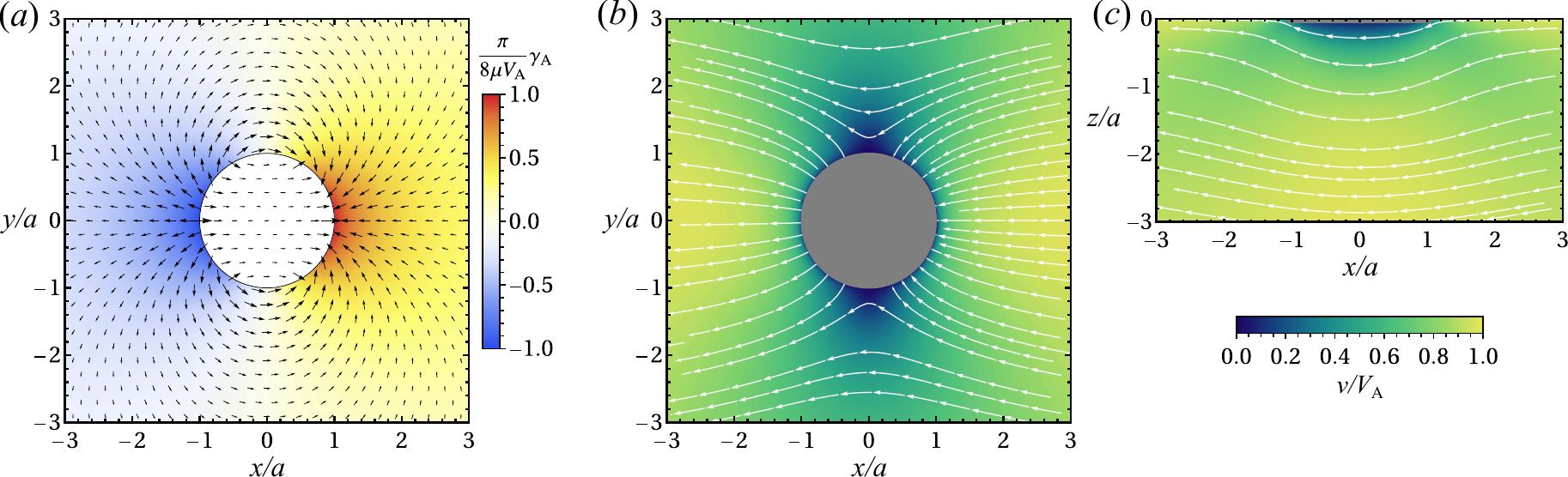}
  \end{center}
  \caption{\label{fig:5}(\textit{a}) The surface tension $\gamma_\mathrm{A}$ of the optimal circular Marangoni surfer (colour scale), relative to the unperturbeed surface. The arrows indicate the force density on the fluid, $\boldsymbol{\nabla}_\mathrm{s}\gamma_\mathrm{A}$ outside the disk and $\bm{f}_\parallel$ underneath it. (\textit{b,c}) The flow field in the co-moving frame in top view (\textit{b}) and side view (\textit{c}).}
\end{figure}

With the drag coefficients of the two passive problems, the respective translational velocities in the passive problems follow from Eq.~\eqref{vFS_vSI} as
\begin{equation}
  \frac{V_\mathrm{FS}}{V_\mathrm{A}} = \frac{R_\mathrm{SI}}{R_\mathrm{SI} - R_\mathrm{FS}}=3 \, , \qquad
  \frac{V_\mathrm{SI}}{V_\mathrm{A}} = \frac{R_\mathrm{FS}}{R_\mathrm{SI} - R_\mathrm{FS}}=2 \, . \label{vFS_vSI_trans}
\end{equation}
The full solution for the optimal active surfer can now be obtained as the superposition of the solution for the passive body in an incompressible surface (Sect.~\ref{sec:3}) with the negative of the solution for the free surface (Sect.~\ref{sec:4}). We thereby replace the body velocity in the former case with $V_\mathrm{SI}=2V_\mathrm{A}$ and in the latter case with  $V_\mathrm{FS}=3V_\mathrm{A}$. 
The relative surface tension of the optimal active surfer (Fig.~\ref{fig:5}a) follows from Eq.~\eqref{eq:gamma}  as $\gamma_\mathrm{A}=-\gamma_\mathrm{SI}$
\begin{equation}
	\gamma_\mathrm{A} = \frac{8\mu V_\mathrm{A}}{\pi} \frac{a}{\rho} \, \cos\theta\,. \label{eq:surf-tension_active}
\end{equation}
The total force on the contact line then equals $8\mu a V_\mathrm{A}$. Interestingly the force is by a factor of $3/2$ larger than the force that would be needed to pull a passive disk with the velocity $V_\mathrm{A}$, as already stated by \citet{Lauga.Davis2012}. Note that the fact that the ratio $3/2$ is the same as the ratio $R_\mathrm{SI}/R_\mathrm{FS}$ is specific to the disk geometry and not generally valid. 

The force density with which the bottom surface of the surfer acts on the fluid can likewise be obtained as ${\vect{f}_\parallel}_\mathrm{FS} - {\vect{f}_\parallel}_\mathrm{SI}$.
The surface integral of this force over the disk equals the force caused by the surface tension, in accordance with the force-free condition on the active surfer. 

Finally, the flow field induced by the optimal Marangoni surfer is obtained as
\begin{subequations}
	\begin{eqnarray}
		\frac{v_\rho}{V_\mathrm{A}} &=& \frac{2}{\pi} \left( C_- + D_- \frac{z}{\rho} \right) \cos\theta \, , \\
		\frac{v_\theta}{V_\mathrm{A}} &=& -\frac{2}{\pi} \left( C_+ + D_+ \frac{z}{\rho} \right) \sin\theta \, , \\
		\frac{v_z}{V_\mathrm{A}} &=& \frac{4}{\pi} \frac{z}{\rho} \, D_1 \cos \theta \, ,
	\end{eqnarray}
\end{subequations}
where we have introduced the abbreviation $C_\pm = C_0 \pm C_2$ and $D_\pm = D_0 \pm D_2$.

At the interface where $z=0$ and $\rho \ge a$, the velocity can be expressed as
\begin{equation}
	\frac{{\vect{v}}_\parallel}{V_\mathrm{A}} =\frac{2}{\pi}\left( C_- \cos\theta \, \hat{\vect{e}}_\rho - C_+ \sin\theta \, \hat{\vect{e}}_\theta \right) \, , 
\end{equation}
with
\begin{equation}
	C_\pm =  \arcsin \left( \frac{a}{\rho} \right) \pm \frac{a \sqrt{\rho^2-a^2}}{\rho^2} .
\end{equation}
The resulting flow field is shown in Fig.~\ref{fig:5}(b,c). The pressure can likewise be written in a closed form as
\begin{equation}
  p(\rho, z=0)= \frac{8\mu V_\mathrm{A}}{\pi\rho}  \frac{a \cos\theta}{\sqrt{\rho^2-a^2}} \, ,
\end{equation}
for $\rho \ge a$ and is zero underneath the surfer.

\section{Discussion}

We have derived a minimum dissipation theorem that gives a lower bound on hydrodynamic dissipation by a Marangoni surfer of arbitrary shape in the limit of low Reynolds and capillary numbers. In our calculation we assumed a planar gas-liquid interface, but the expansion to any other interface form, for example, taking into account the meniscus around the contact line, is straightforward as long as the shape is not influenced by the flow. Likewise, the problem can be solved with additional confinement for the liquid phase where an interesting  reversal of swimming direction can occur \citep{Vandadi.Masoud2017}. Another generalisation would be to include the rotational motion of the surfer. In all those cases the only difficulty is to calculate the drag coefficients of the corresponding passive problems.

As a specific example, we calculated the efficiency limit for a surfer taking the form of a thin circular disk. The obtained value $\eta_\mathrm{L}=1/3$ is significantly higher than any other realistic swimmer designs. Interestingly, \citet{Lauga.Davis2012} have already discussed exactly the same solution that we have now proven to be hydrodynamically optimal. Their approach is, however, very different from ours as it is based on the solution of the Laplace equation for the diffusion of surfactants around the surfer in the limit of low Péclet numbers. They have shown that only the first angular mode, corresponding to a dipolar source of surfactants, contributes to propulsion. We have now shown that this mode coincides with the solution of the flow optimisation problem. The two solutions are not entirely comparable, because at low Péclet numbers surfactant diffusion makes the dominant contribution to the total dissipation. On the other hand, the finding that surfactant distributions resulting from diffusion lead to hydrodynamically optimal solutions also has a broader validity, beyond disk-shaped bodies. From the structure of our Green's functions one can see that the optimal distribution of surface tension indeed satisfies a 2D Laplace equation around any flat surfer. We expect, however, that this is not the case for general surfers with a finite vertical extension.  

While we are now able to determine the solution with the minimal hydrodynamic dissipation, losses of free energy also take place due to surfactant diffusion. If the surface tension is related to the surface density of insoluble surfactants $c$ as $\gamma=\gamma^0 - K c$ and the dynamics of the latter is determined by the advection-diffusion equation \citep{Schwartz.Roy2001}
\begin{equation}
  \partial_t c = - \boldsymbol\nabla_\mathrm{s}\cdot ( \boldsymbol{v} c - D\boldsymbol{\nabla}_\mathrm{s} c)\,,
\end{equation}
the rate of free energy loss due to diffusion can be written as
\begin{equation}
  P_\mathrm{D}=  K D \int_\mathcal I \dd S\, \frac{(\boldsymbol{\nabla}_\mathrm{s} c)^2}{c}\;.
\end{equation}
The rate of work done on the fluid (\ref{eq:dissipation2}), on the other hand, is $P=  K  \int_\mathcal I \dd S\, c  \boldsymbol{\nabla}_\mathrm{s} \cdot \boldsymbol{v}$. Their ratio is determined by the Péclet number, defined as
\begin{equation}
  \label{eq:peclet}
  Pe=\frac{v}{D a}\,,
\end{equation}
if we assume that the derivatives of $c$ and $\mathbf{v}$ are determined by the characteristic length scale $a$.
Our theorem therefore gives the total dissipation in the limit of high Péclet numbers. In the general case, it applies to the hydrodynamic dissipation only. 
At the same time, diffusivity is also needed if, for example, the surfer only emits surfactants at its perimeter, but not elsewhere. This limitation does not apply to all surfers -- a counterexample is given by beetles that can secret surfactants from their tail, at a distance from other contact lines \citep{Lang.Dettner2012}. Furthermore, our study cannot take into account the energetic cost of synthesising or collecting surfactants. This cost could be significant in the total energy balance of propulsion. Overall, finding optimal solutions for the combined problem including hydrodynamics and diffusion of surfactants remains a challenge for future investigations.

\acknowledgements{The authors thank the anonymous reviewer of Ref.~\onlinecite{DaddiMoussaIder.Vilfan2023} for the suggestion to derive a minimum dissipation theorem for Marangoni surfers.}

\appendix
\section{Derivation of the passive minimum dissipation theorem}
\label{sec:helmholtz}

The Helmholtz minimum dissipation theorem states that among all incompressible flows that satisfy a prescribed fixed-velocity boundary condition, the Stokes flow has the smallest dissipation. It can be generalized to many other boundary conditions. For example, among all flows with zero normal velocity at the surface of a body, the flow with vanishing tangential tractions at the interface (perfect-slip boundary condition) has the minimum dissipation \citep{Nasouri.Golestanian2021}. More generally, at a fluid-fluid interface, dissipation is minimal for the flow with stress continuity \citep{DaddiMoussaIder.Vilfan2023}. In the following, we show that among all flows around a body submerged in a planar gas-liquid interface and moving with velocity $\boldsymbol{V}$, the free-surface flow has the smallest dissipation. The derivation is adapted from the proof of the original Helmholtz minimum dissipation theorem as formulated in the textbook by \citet{guazzelli2009} and follows a similar line as our previous generalizations \citep{Nasouri.Golestanian2021,DaddiMoussaIder.Vilfan2023}. 

We consider the unperturbed flow with velocity $\boldsymbol{v}$ and boundary conditions as depicted in Fig.~\ref{fig:2}, i.e., $\boldsymbol{v}=\boldsymbol{V}$ at the swimmer surface $\mathcal{S}$ and $\boldsymbol{v}\cdot \hat{\boldsymbol{e}}_z=0$ at the interface $\mathcal{I}$. The total dissipation in the fluid domain is 
	\begin{equation}
		P=2\mu \int_\mathcal{V}\dd V \, \boldsymbol{E}:\boldsymbol{E}\,.
	\end{equation}
We now perturb the velocity field by $\boldsymbol{v}'$. The perturbed flow also needs to satisfy the boundary conditions, therefore $\boldsymbol{v}'=\boldsymbol{0}$ at $\mathcal{S}$ and $\boldsymbol{v}'\cdot\hat{\boldsymbol{e}}_z=0$ at  $\mathcal{I}$. The perturbation to the strain rate is $\boldsymbol{E}' = \left( \boldsymbol{\nabla} \boldsymbol{v}' + (\boldsymbol{\nabla} \boldsymbol{v}')^\top \right)/2$. Its effect on the dissipation follows as 
\begin{equation}
  \Delta P=2\mu \int_\mathcal{V} \dd V \left[ \left(\boldsymbol{E}'+\boldsymbol{E}\right):\left(\boldsymbol{E}'+\boldsymbol{E}\right)-\boldsymbol{E}:\boldsymbol{E}\right] = 2\mu \int_\mathcal{V}\dd V \, \boldsymbol{E}':\boldsymbol{E}'
  + 4\mu \int_\mathcal{V}\dd V \, \boldsymbol{E}':\boldsymbol{E}\,.
  \label{eq:deltap1}
\end{equation}
Both tensors $\boldsymbol{E}'$ and $\boldsymbol{E}$ are traceless and symmetric. We can therefore write $2\mu \boldsymbol{E}':\boldsymbol{E}=\boldsymbol{\nabla}\boldsymbol{v}':\boldsymbol{\sigma}$, using the stress tensor $\boldsymbol{\sigma}=-p\boldsymbol{I}+2\mu\boldsymbol{E}$. In the absence of volume forces, the divergence of the stress tensor vanishes, $\boldsymbol{\nabla}\cdot\boldsymbol{\sigma}=\boldsymbol{0}$ and we can write $\boldsymbol{\nabla}\boldsymbol{v}':\boldsymbol{\sigma}=\boldsymbol{\nabla}\cdot\left(\boldsymbol{v}'\cdot\boldsymbol{\sigma}\right)$. Finally, by applying the divergence theorem, it follows $4\mu \int_\mathcal{V}\dd V \, \boldsymbol{E}':\boldsymbol{E}=-2\int_\mathcal{S} \dd S\, \boldsymbol{v}' \cdot \boldsymbol{\sigma}\cdot \boldsymbol{n}+ 2\int_\mathcal{I} \dd S\, \boldsymbol{v}' \cdot \boldsymbol{\sigma}\cdot \hat{\boldsymbol{e}}_z$. The first integral vanishes because $\boldsymbol{v}'=\boldsymbol{0}$ at the swimmer surface $\mathcal{S}$. The second vanishes because there are no tangential tractions at the interface, $\boldsymbol{\sigma}\cdot \hat{\boldsymbol{e}}_z \parallel \hat{\boldsymbol{e}}_z$ while $\boldsymbol{v}' \cdot \hat{\boldsymbol{e}}_z=0$. Together, this shows that the second term in Eq.~(\ref{eq:deltap1}) is always zero. The remaining term, $\int \dd V \, \boldsymbol{E}':\boldsymbol{E}'$, is positive for any non-vanishing perturbation. It follows that the minimum dissipated power under the requested boundary conditions is achieved exactly when $\boldsymbol{E}'=\boldsymbol{0}$. We have thus proven that any deviation from the force-free flow around a moving body embedded in a planar interface leads to increased dissipation.

\bigskip

\bibliography{reference}
 
\end{document}